\documentclass[twocolumn]{article}
\usepackage[a4paper, total={6in, 8in}]{geometry}

\usepackage[backend=biber, style=trad-abbrv]{biblatex}
\usepackage{amsfonts, amsthm}
\usepackage{algorithmic}
\usepackage{textcomp}
\usepackage{wrapfig}
\usepackage{tcolorbox}
\usepackage{url}
\theoremstyle{definition}
\newtheorem{definition}{Definition}
\usepackage[hidelinks]{hyperref}

\usepackage[noabbrev,nameinlink]{cleveref} 
\usepackage{svg}
\svgsetup{inkscapelatex=false}
\bibliography{picarx}

\crefformat{footnote}{\protect#2\footnotemark[#1]\protect#3\protect}

\usepackage{xspace}
\usepackage{eurosym}
\usepackage{dblfloatfix}

\newcommand{\ds}{digital shadow\xspace}
\newcommand{\dt}{digital twin\xspace}
\newcommand{\dts}{digital twins\xspace}
\newcommand{\pt}{physical twin\xspace}
\newcommand{\pts}{physical twins\xspace}

\newcommand{\dtp}{digital twin prototype\xspace}
\newcommand{\dtps}{digital twin prototypes\xspace}
\newcommand{\gazebo}{GAZEBO\xspace}

\begin{document}

\title{Toward Reproducibility of Digital Twin Research:\\Exemplified with the PiCar-X}
%
%

\author{Alexander Barbie \&\ Wilhelm Hasselbring\\
Software Engineering Group, Kiel University, Germany}

\maketitle

\begin{abstract}
Digital twins are becoming increasingly relevant in the Industrial Internet of Things and Industry 4.0, enhancing the capabilities and quality of various applications. However, the concept of \dts lacks a unified definition and faces validation challenges, partly due to the scarcity of reproducible modules or source codes in existing studies. While many applications are described in case studies, they often lack detailed, re-usable specifications for researchers and engineers.
In previous research, we defined and formalized the \dt concept. This paper presents a reproducible laboratory experiment that demonstrates various \dt concepts. Our formalized concept encompasses the \pt, the digital model, the digital template, the digital thread, the digital shadow, the \dt, and the \dtp. We illustrate this series of concepts by using a PiCar-X, showcasing the progression from a \pt to its \dtp. The entire code base is published as open source, and for each concept, Docker-compose files are provided to facilitate independent exploration, understanding, and extension.
\end{abstract}


\section{Introduction}
The field of embedded software engineering is undergoing a significant transformation with the introduction of the Internet-of-Things (IoT) and cyber-physical systems (CPS) paradigms. These paradigms offer numerous advantages for applications in the context of Industry 4.0, but also have disadvantages and challenges. These challenges have been described in various publications~\cite{Jazdi2014,dtdef-koulamas-cps,cpssimulationmanufacturing}.

Embedded software systems are no longer self-contained and can consist of numerous interconnected devices. The mandatory components are a (central) processing node and compatibility among heterogeneous devices. The task at hand is to ensure their interoperability. The increased complexity of these embedded software systems poses a challenge, arising from the continuous information exchange between devices and the implementation of failure prevention mechanisms. A significant challenge lies in managing this software complexity, including the infrastructure, with a server that oversees a potentially vast network of interconnected devices \cite{Mittal2019}. 

For decades, enhanced tools, methods, and technologies addressing different challenges, e.g. proper security measures and complexity of the software \cite{iotchallenges2016}, were introduced. One of the emerging technologies, is the \dt paradigm. Similar to the NASA's ``Twin'' approach during the Apollo Missions in the 1960, where always two identical space capsules were build and one remained on earth to allow experiments on it before changes were made by Astronauts in space, the \dt is a mirror of a real physical object \cite{dtdef-barbie}. The advantage over a second \pt, is the price. A virtual object can be instantiated many times, does not need the same amount of maintenance, and enables a better collaboration among engineers located at different places around the world. However, the \dt is still a green field and various definitions and applications exist without a consensus on one definition. We discussed advantages and disadvantages of different \dt definitions in \textcite{dtdef-barbie} and, in addition, fully formalized the \dt concept using the Object-Z notation~\cite{smith2012object}. The formalized concept was demonstrated in a real-life mission, where an underwater network of ocean observation systems was established in the Baltic Sea~\cite{demomission}. Additionally, the \dtp approach was utilized in the development of a smart farming application \cite{silagecontrol}. 

Validation of research results and the reproducibility of experiments is part of good scientific practice \cite{empiricalstandards}, in particular for benchmarking~\cite{EASE2021}. Yet reproducing, for instance, an ocean observation network requires enormous resources for other researchers, not only to reach the same location to validate the gathered data, but also the used hardware is quite specific for ocean science and expensive \cite{demomission}. Research on the reproducibility of experiments in the software engineering community indicates that there is concern about the quality of this kind of validation in the community~\cite{Shepperdreproduceability}. One reason is the effort needed to produce validation studies due to poor support for reproducibility~\cite{Madeyski2017}. 
Thus, we see the need for an affordable laboratory example for independent replication \cite{basili2005experimental} of our \dt concept. Especially, since there are only few reproducible examples of \dts, researchers and engineers could use to explore the \dt concept. \textcite{Fogli2023} describe a simple experiment for chaos testing the resilience of \dts, however, the presented \dt would be in a very early conceptual stage. By the time this paper was published, we did not find a reproducible experiment to exemplify the basic implementations of the various concepts from \pt to \dt.

The open science agenda holds that science advances faster when we can build on existing results. Replicability and reproducibility are the ultimate standards by which scientific claims are judged~\cite{Peng2011}. However, the respective terminology is often confused \cite{plesser_reproducibility_2018}. We follow the current ACM terminology for Artifact Review and Badging of repeatability, reproducibility and replicability.\footnote{\url{https://www.acm.org/publications/policies/artifact-review-and-badging-current}} Please note that the ACM swapped the terms `reproducibility' and `replication' from Version~1.0 to Version~1.1 as a result of discussions with the National Information Standards Organization (NISO).\footnote{\url{https://www.acm.org/publications/badging-terms}} We follow the current Version~1.1. A study is \textit{repeated}, if the same team obtains the same or similar results by applying the same experimental setup. For \textit{reproducibility}, a different team obtains the results with the same experimental setup. If the same results are achieved by a different team that applied a different experimental setup, the study is \textit{replicable}.\footnote{See also \url{https://github.com/acmsigsoft/EmpiricalStandards/blob/master/docs/standards/Replication.md}} Our goal is enabling the reproducibility (and extension) of digital twin experiments by other research groups. 

In this paper, the various elements of our \dt concept \cite{dtdef-barbie} are showcased using the PiCar-X by SunFounder.\footnote{\label{fn:picarx}\url{https://www.sunfounder.com/products/picar-x}} This approach systematically demonstrates each concept, illustrating the theory with practical examples. We begin with an overview of the PiCar-X, followed by a step-by-step illustration of various concepts: starting with the \pt (\Cref{subsec:PhysicalTwin}), moving through the digital model (\Cref{subsec:DigitalModel}), digital template (\Cref{subsec:digitaltemplate}), digital thread (\Cref{subsec:digitalthread}), digital shadow (\Cref{subsec:digitalshadow}), \dt (\Cref{subsec:digitaltwin}) and finally, the \dtp (\Cref{subsec:dtp}). Each concept is supported by a Docker-compose file to demonstrate the differences, and the entire code base is made available as open source \cite{abarbiegithub}. In \Cref{sec:experiment}, a practical experiment that readers can replicate, is included. \Cref{sec:Conclusion} concludes the paper.

\section{Exemplifying of the Digital Twin Concept}
\begin{figure}[ht]
    \centering
    \includegraphics[width=0.4\textwidth]{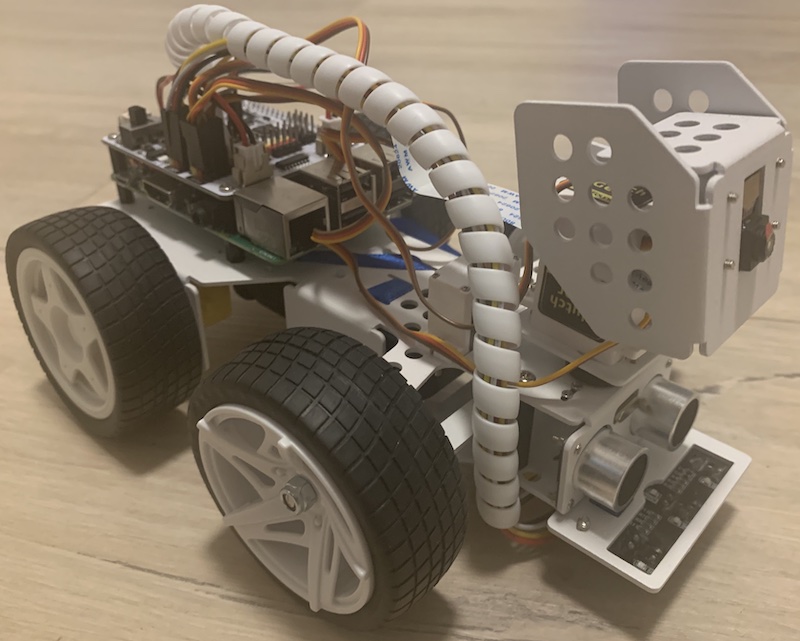} 
    \caption{The \pt of the PiCar-X.}
    \label{fig:picarx}
\end{figure}

We present a reproducible lab experiment: the SunFounder PiCar-X, shown in \Cref{fig:picarx}, for independent replication of the \dt approach by engineers and other researchers. The PiCar-X kit\cref{fn:picarx} costs around \EUR{140}, and the RaspberryPi 3B+/4 has additional costs of around \EUR{50}. All in all, the hardware costs for the PiCar-X are around \EUR{200}.

The PiCar-X is moved by two direct current (DC) geared motors, one for each rear wheel. The two front wheels are steered by a servo motor. The two remaining servo motors turn the camera left/right and up/down. An ultrasonic sensor at the front is used for collision avoidance. The gray-scale sensor is placed on the ultrasonic module and is used for line following. The two main interfaces are GPIO and I2C. Notice that SunFounder calls the GPIO pins connected to DC and servo motors pulse width modulation (PWM) \emph{pins}. These PWM pins are I2C-controlled. The motors use the GPIO pins to drive forward (0) or backward (1). The speed of the DC motors and the angle of the servo motor are set via an I2C interface. A setup with two back wheels that accelerate the car and steering by turning only the two front wheels is a classic Ackermann steering setup known in common cars \cite{Veneri2020}.
In this paper, we focus on the PiCar's driving capabilities. The remaining sensors were not implemented by the time this paper is written.

The source code is published on GitHub~\cite{abarbiegithub} and is based on the ARCHES Digital Twin Framework (ADTF)~\cite{ADTF}, both are published open source.
Although originated from the research on ocean observations systems~\cite{demomission}, the implemented framework is independent from the domain and hence, can be reused anywhere, with one restriction, the application does not have to meet hard real-time requirements. The ADTF is built upon the open source middleware Robot Operating System (ROS or ROS1)~\cite{rospaper}, which does not meet real-time requirements.
We use three principles from the ADTF~\cite{ADTF}:
\begin{itemize}
    \item Data/status messages are automatically sent from the \pt to the \dt.
    \item Command messages are automatically sent from the \dt to the \pt.
    \item Nodes are deployed in Docker containers.
\end{itemize}

While containerization, e.g. with Docker, is already state-of-the-art in cloud-based services and many other applications, the adaption for industrial applications lags behind. One reason is the resulting communication overhead between containers, which, to date, does not reliably meet real-time requirements in industrial contexts. Since ROS also does not support hard real-time applications, we ignore this handicap of Docker. Instead, we exploit an advantage of containerization with Docker. Without containerization, developers start all components as processes in their development environments \cite{demomission}. This can be laborious, especially, if one has to keep the dependencies and configurations up-to-date. Even small differences between two development environments can later result in bug fixing. Containerization avoids this problem since the base container has always the same (clean) setup. Another advantage is that platform-independent software development is possible.

\newpage 

\subsection{The Physical Twin}
\label{subsec:PhysicalTwin}
With the increasing demand for context-aware, autonomous, and adaptive robotic systems~\cite{DTroadmap}, the embedded software community must adopt more advanced software engineering methods. As a result, the development processes for these systems must evolve. In future workflows, embedded software systems will become the centerpiece of a product. To achieve this, the community needs to shift from expert-centric tools~\cite{DTroadmap} to modular systems that empower domain experts to contribute to specific parts of the system~\cite{dtdef-barbie}. For instance, in the context of an autonomous car, the vehicle integrates multiple autonomous subsystems, such as sensors, navigation, communication, and control systems, each with distinct functionalities. These systems collaborate to enable the car to perceive its environment, make decisions, and act in real-time. This integration of software modules and physical components allows the car to operate autonomously in complex and dynamic environments, forming a system of systems~\cite{SoS-Boardman}.

While the PiCar-X is a less complex embedded system, we use it to exemplify our \dt concept, which begins with the \pt~\cite{dtdef-barbie}:

\begin{tcolorbox}
\begin{definition}[Physical Twin]\label{def:pt}
A \pt is a real-world physical System-of-Systems or product. It comprises sensing or actuation capabilities driven by embedded software.
\end{definition}
\end{tcolorbox}

While SunFounder offers a Python implementation for the PiCar-X, we found it inadequate and opted to reimplement all components from scratch. This approach allowed us to adhere to ROS's ``thin'' paradigm \cite{rospaper}, resulting in one package dedicated solely to PiCar-X functionalities and another ROS package that incorporates the first as a dependency, creating nodes around it. Furthermore, we embraced the clean architecture principles for embedded software by \textcite{cleanarchitecturebook}, focusing on drivers that manage hardware interactions and relay data to other processes.

\begin{figure*}[ht]
    \centering
    \includegraphics[width=.85\textwidth]{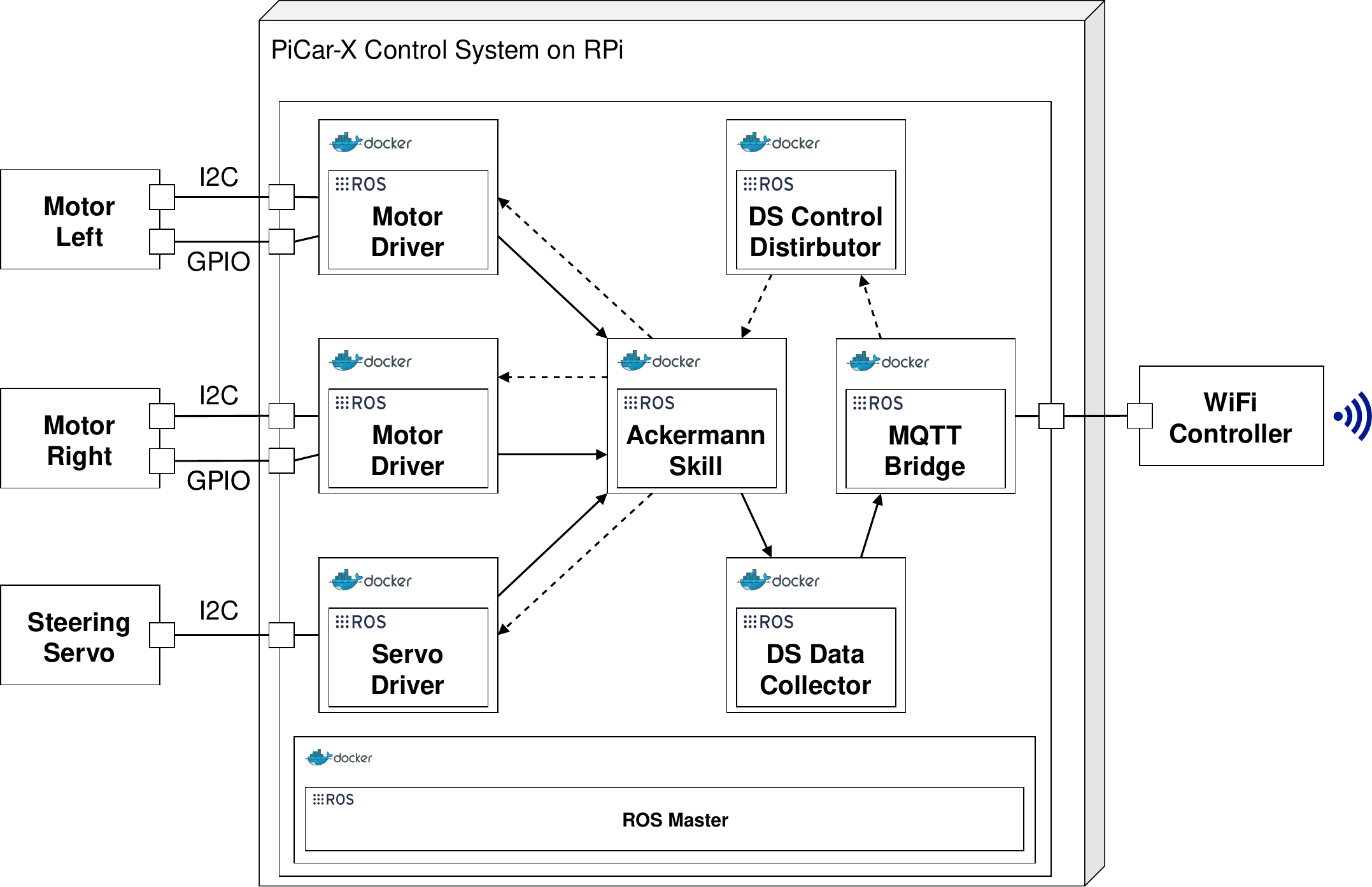}
    \caption{Software running the PiCar-X \pt.}
    \label{fig:picarxpt}
\end{figure*}

The base container employs ROS Noetic on Ubuntu 20.04, forming the foundation of the PiCar-X's software setup, see \Cref{fig:picarxpt}. The servo motor driver implementation for the PiCar-X is straightforward. The servo motor, which controls the car's steering, operates within a range of $\alpha \in [0, 180]$ degrees. The angle is controlled by writing a pulse width to an I2C register. For straight forward motion, the servo motor is set to $90^{\circ}$ degrees. Angles less than $90^{\circ}$ result in a left turn, while those greater than $90^{\circ}$ cause a right turn. It is evident that the wheels cannot be turned to $0^{\circ}$ or $180^{\circ}$; the construction of the car allows for a maximal angle of $\alpha_{possible} \in (50, 140)$ or $\alpha_{left} \in (-40, 0]$ and  $\alpha_{right} \in [0, 40)$.

The control mechanism for each DC motor in the PiCar-X involves a combination of a single GPIO pin and an I2C register. The GPIO pin's state dictates the motor's direction: a value of 1 for forward motion, and 0 for backward. Since the motors are identical and each drives a separate wheel, one motor is inverted to ensure coordinated movement on both sides. Thus, while the left motor moves forward when its GPIO pin is set to 1, the right motor moves forward when its GPIO pin is at 0. The motors' speed is regulated by the pulse width written to their respective I2C registers.
In this setup, the PiCar-X uses the \emph{Ackermann Steering Skill}. A \emph{Skill} is an interface provided by the ADTF, which refers to nodes with publishers or subscribers that can synchronize messages with the physical or \dt. The \emph{Ackermann Steering Skill} node processes incoming commands, sending speed and direction data to motor drivers, and angle information to the servo driver.

Implementation of other sensors like grey-scale, avoidance detection, and camera is not covered in this example, so far. However, since the I2C chip interactions are already established for the DC motor driver and emulator, other researchers are encouraged to experiment with these functionalities. Communication with the \dt is conducted via WiFi using the MQTT protocol.

\newpage

\subsection{The Digital Model}
\label{subsec:DigitalModel}
The first ``digital'' part of the \dt concept is the \textit{Digital Model}~\cite{dtdef-barbie}:
\begin{tcolorbox}
\begin{definition}[Digital Model]
A digital model describes an object, a process, or a complex aggregation. The description is either a mathematical or a computer-aided design (CAD).
\end{definition}
\end{tcolorbox}

\begin{figure}[ht]
    \centering
    \includegraphics[width=0.45\textwidth]{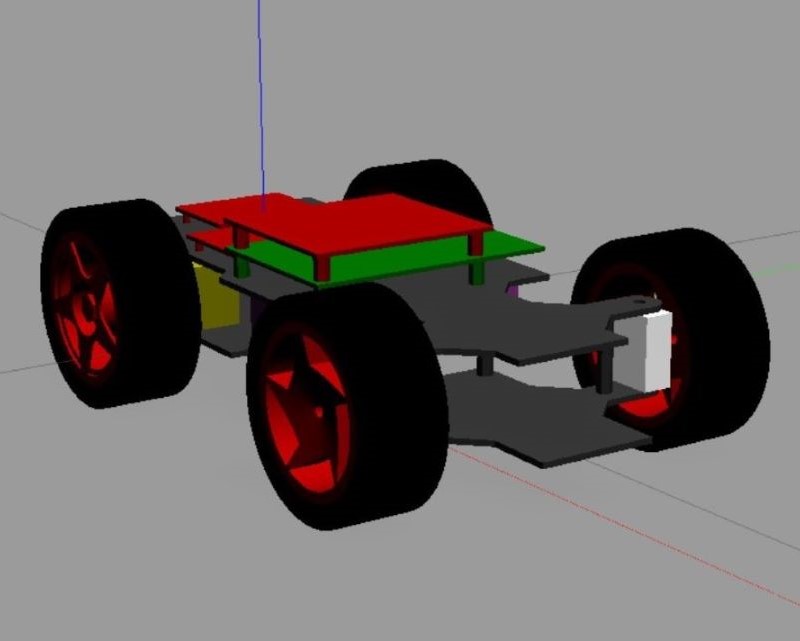}
    \caption{The digital model of the PiCar-X used in the \gazebo simulation.}
    \label{fig:picarxcad}
\end{figure}

In our formalization \cite{dtdef-barbie}, the digital model is represented by a software state machine model. This PiCar-X example demonstrates the approach with a CAD model in a \gazebo simulation, an open-source tool integrated with ROS~\cite{gazebopaper}, as digital model. Lacking official CAD files for the PiCar-X, we utilized a simplified CAD model of an older SunFounder PiCar version, see \Cref{fig:picarxcad}, available under Apache 2.0 license on GitHub.\footnote{\url{https://github.com/Theosakamg/PiCar_Hardware}} This model, consisting of just the frame and wheels, closely mirrors the original PiCar-X's key dimensions like wheelbase and track, crucial for an accurate Ackermann steering simulation. However, the original PiCar-X's steering mechanism, operated by a steering bar to achieve Ackermann angles, could not be replicated in \gazebo. Instead, we approximate the Ackermann steering angles for both front wheels based on established methodologies~\cite{Veneri2020}.

\begin{figure}[ht]
    \centering
    \includegraphics[width=0.45\textwidth]{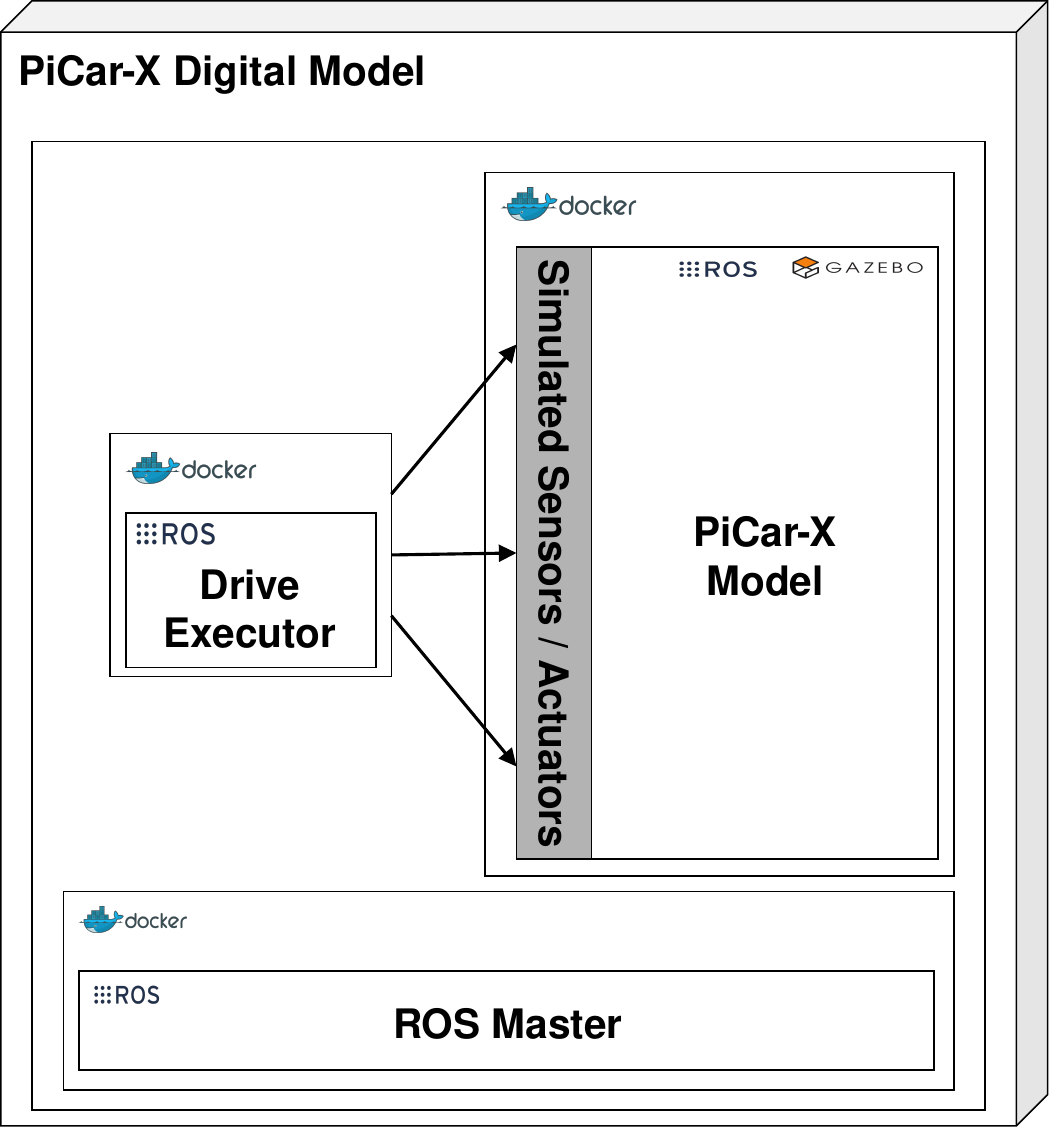} 
    \caption{The digital model of the PiCar-X, including the software component and the CAD model in a \gazebo simulation.}
    \label{fig:picarxdm}
\end{figure}

The digital model's setup, as illustrated in \Cref{fig:picarxdm}, incorporates the CAD model into the \gazebo simulation. The `\textit{ros\_control}' package is used to simulate the sensors and actuators, which offers topics for manipulating the model's joints for steering and wheel movement. In this setup, four joints need manipulation, including approximating the steering angles. Direct command-line control in ROS could introduce delays, and since the PiCar-X operates with pulse widths rather than angles and speed values, the digital model includes a node that translates pulse widths into angles. This node receives a \textit{DriveStatus} message, converts the pulse widths for the motors and steering, and then publishes the calculated values to the joints. Both the ROS node and the \gazebo simulation are connected to the same ROS Master, ensuring seamless integration.

This represents a basic setup for the digital model. While it can be expanded to test additional functions or skills, at this stage, the digital model is not yet connected to the \pt. Consequently, any modifications made on the \pt must be manually replicated on the digital model, and the same applies in reverse.

\subsection{The Digital Template}\label{subsec:digitaltemplate}
Bundling the PiCar's \pt software, its digital model, blue prints, and other documents, forms the \textit{Digital Template}. The digital template is defined as follows \cite{dtdef-barbie}:
\begin{tcolorbox}
\begin{definition}[Digital Template]
A digital template serves as a framework that can be tailored or populated with specific information to generate the \pt. It encompasses the software operating the \pt, its digital model, and all the essential information needed for constructing and sustaining the \pt, such as blueprints, bills of materials, technical manuals, and similar documentation.
\end{definition}
\end{tcolorbox}

The following components can be identified for the digital template:
\begin{itemize}
    \item the \pt software from GitHub \cite{abarbiegithub};
    \item the digital model from GitHub \cite{abarbiegithub};
    \item the documentation including the list of materials, the robot HAT blue prints, etc.;\footnote{\url{https://docs.sunfounder.com/projects/picar-x/en/latest/index.html}}
    \item the Raspberry Pi documentation, including mechanical drawings and schematics.\footnote{\url{https://www.raspberrypi.com/documentation/computers/raspberry-pi.html}}
\end{itemize}

The current configuration lacks the blueprints for key hardware components such as servo and DC motors. Ideally, a complete digital template would encompass all necessary assets to replicate the \pt from scratch. However, this scenario is often impractical, as most machines are composed of various components, and it is rare for a single company to manufacture every part. 

\subsection{The Digital Thread}\label{subsec:digitalthread}
Physical twin and \dt are connected via a \textit{Digital Thread}, which is defined as \cite{dtdef-barbie}:
\begin{tcolorbox}
\begin{definition}[Digital Thread]\label{def:digitalthread}
The digital thread refers to the communication framework that allows a connected data flow and integrated view of the \pt's data 
and operations throughout its lifecycle.
\end{definition}
\end{tcolorbox}

The first part of the digital thread was already shown for the digital model, where the \gazebo simulation is utilized to visualize the CAD model. The ADTF facilitates communication through two mechanisms: decorated publishers/subscribers for command or data/status messages and an automated exchange system. Commands are automatically relayed to the \pt, and data/status messages from the \pt are transmitted to a server via MQTT over wireless LAN. Three ROS nodes on the \pt handle this data exchange. Additionally, these nodes serialize and deserialize ROS messages to optimize data transmission, using Apache Avro\footnote{\label{fn:avro}\url{https://avro.apache.org/}} for efficient binary message serialization. This setup ensures efficient and accurate message exchange between the digital and physical components.

Three ROS nodes (\textit{DS Data Collector}, \textit{MQTT Bridge}, \textit{DS Control Distributor}, see \Cref{fig:dtpsetup}) form the digital thread on the \pt. 
The \textit{DS Data Collector} handles all data/status messages, forwarding them to the digital shadow or twin via the \textit{MQTT Bridge}. Meanwhile, the \textit{MQTT Bridge} receives incoming messages, which the \textit{DS Control Distributor} then allocates to appropriate topics. On the opposite side, a \textit{DS Data Distributor} manages data/status messages from the \pt, distributing them to corresponding topics. Additionally, the \textit{DS Control Collector} gathers commands and transmits them to the \pt through the \textit{MQTT Broker}.

Besides their primary role in collecting and distributing ROS messages, these nodes have a secondary function: serializing and deserializing ROS messages for efficient communication. Since ROS message exchange directly would require the \pt to be constantly connected to a server running a ROS Master, this approach is not feasible. ROS messages are typically in an inline-YAML format, which is readable but bulky. To streamline this, messages are converted into a compact binary format using Apache Avro\cref{fn:avro}, and then reverted back to ROS messages upon receipt. This method reduces the data size, optimizing message transmission between the \pt and its digital counterpart.

For data storage in this setup, ROS's built-in feature for recording messages into ROS bag files can be used. While a database like MongoDB could also be employed for this purpose, it is not essential for the scope of this example.


\subsection{The Digital Shadow}\label{subsec:digitalshadow}
The \textit{Digital Shadow} is what is often presented as \dt~\cite{Brauner2022, Dalibor2022, Lehner2022}. However, a digital shadow is defined as follows \cite{dtdef-barbie}:

\begin{figure*}[b]
    \centering
    \includegraphics[width=.8\textwidth]{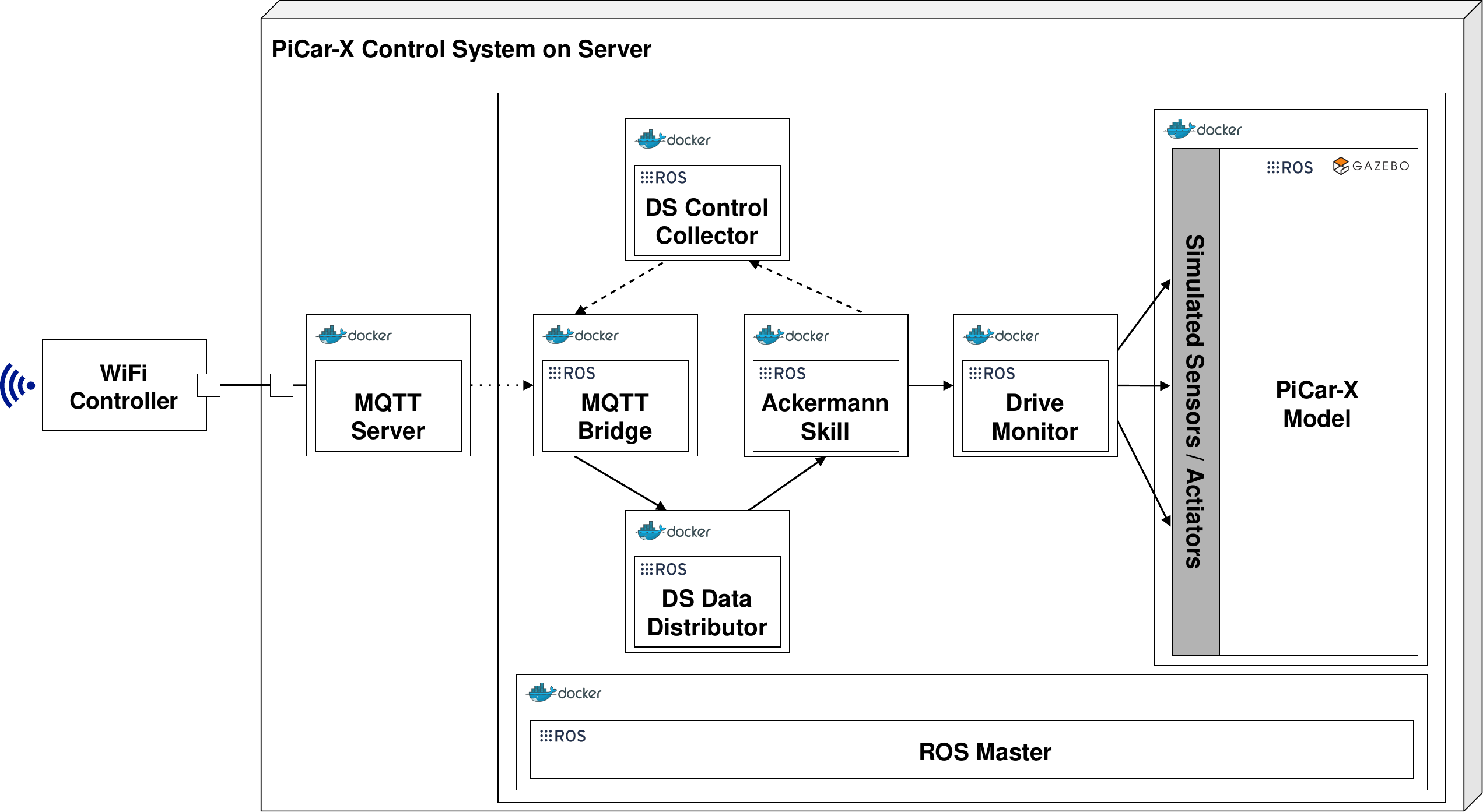} 
    \caption{The digital shadow of the PiCar-X. The communication from \pt to the digital shadow is fully automated.}
    \label{fig:picarxds}
\end{figure*}

\begin{tcolorbox}
\begin{definition}[Digital Shadow]
A digital shadow is the sum of all the data that are gathered by an embedded system from sensing, processing, or actuating. The connection from a \pt to its digital shadow is automated. Changes on the \pt are reflected to the digital shadow automatically. Vice versa, the digital shadow does not change the state of the \pt.
\end{definition}
\end{tcolorbox}

In this setup, monitoring the state of the \pt aligns with CPS concepts, with a distinctive focus on the digital model as a central, continuously updated element reflecting the \pt's state.
In the PiCar-X example, a different approach from the MAPE-K reference model used in our formalization \cite{dtdef-barbie} is adopted. While some kind of self-adapting control loop is common for developing \dts in an Industry 4.0 context \cite{Bibow2020, Michael2024}, its downside is the additional development effort. 
For instance, in a MAPE-K approach, the digital shadow is often developed from scratch and independent from the \pt's embedded software. 
The difference in the control logic between the \pt and digital shadow can impact the fidelity of the digital shadow \cite{Dalibor2022} and later the \dt.

By integrating the essential ADTF~\cite{ADTF} components and the Ackermann skill into the digital shadow, and reusing components from the \pt, the presented approach aims to reduce possible integration errors when incorporating the digital model into the digital shadow. \textcite{Heithoff2023} describe this approach to enhance the sustainability of the embedded control software.
The Ackermann skill consolidates status messages from the \pt's drivers into a \textit{DriveStatus} event, which contains motor pulse widths. 
A separate node, the \textit{Drive Monitor}, interprets these pulse widths as angles and speeds, updating the digital model in the \gazebo simulation. 
This entire process is automated, with no feedback loop to the \pt yet established as illustrated in \Cref{fig:picarxds}.

\subsection{The Digital Twin}\label{subsec:digitaltwin}
A \textit{Digital Twin} can be seen as an extension of the \ds and is defined as \cite{dtdef-barbie}:

\begin{tcolorbox}
\begin{definition}[Digital Twin]\label{def:dtdef-saracco}
A \dt is a digital model of a real entity, the \pt. It is both a digital shadow reflecting the status/operation of its \pt, and a digital thread, recording the evolution of the \pt over time. The \dt is connected to the \pt over the entire life cycle for automated bidirectional data exchange, i.e. changes made to the \dt lead to adapted behavior of the \pt and vice-versa.
\end{definition}
\end{tcolorbox}

\begin{figure*}[t]
    \centering
    \includegraphics[width=.8\textwidth]{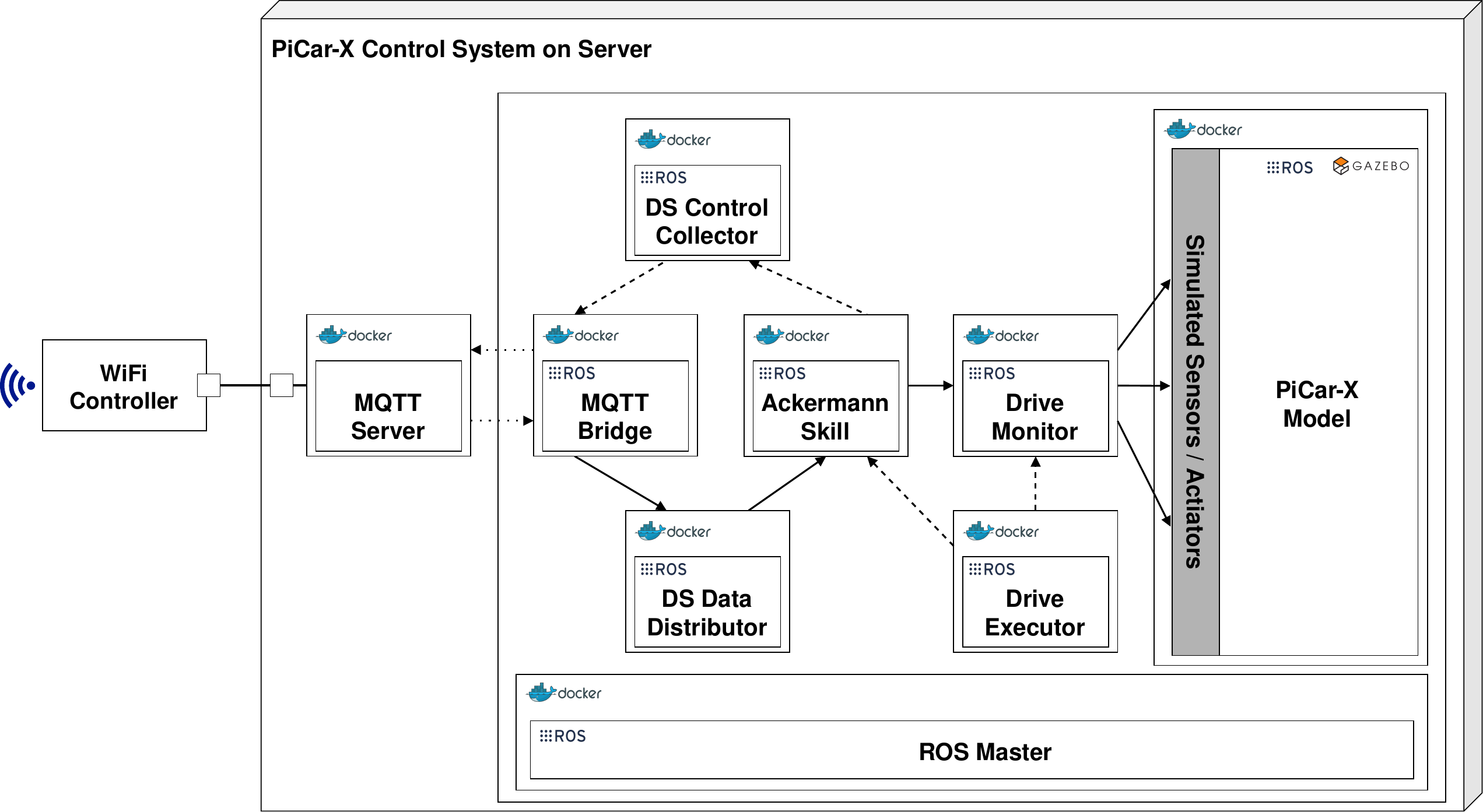} 
    \caption{The \dt of a PiCar-X. The communication from \pt to the \dt and vice versa is fully automated.}
    \label{fig:picarxdt}
\end{figure*}

The primary distinction between a digital shadow and a \dt lies in the nature of their communication with the \pt. In a \dt setup, communication is bidirectional: any changes made to the \dt are automatically replicated in the \pt. 
This is achieved by modifying the \textit{Driver Monitor} node in \Cref{fig:picarxds}, allowing data from the \pt to continue updating the digital model, while the introduction of a 
\textit{Drive Executor} node facilitates sending commands that affect both the digital and \pts simultaneously, see \Cref{fig:picarxdt}. This setup ensures that any alterations in the \dt are mirrored in the physical counterpart.

Reusing the \textit{Ackermann Steering Skill} in the \dt offers an advantage over implementing a separate MAPE-K control loop, particularly in managing steering angles in the digital model. The skill prevents under-steering for angles exceeding 20 degrees, a challenge identified in the digital model due to the Ackermann steering angle approximation. For larger steering angles, the digital model would have a larger turning radius than the real PiCar and hence, the \pt and the \dt would differ while moving. Implementing a separate filtering logic in a MAPE-K process for steering-angle input might introduce additional failure points, as it would require duplication of the filter on both the physical and \dts, rather than reusing the \pt's existing logic.

\subsection{The Digital Twin Prototype}\label{subsec:dtp}
The digital shadow and \dt setups highlight ongoing challenges in accurately mirroring the state of a \pt. A significant issue is the exclusion of many software components that operate the \pt, particularly the device drivers due to their need for hardware interfaces. The \gazebo simulation, and simulations in general, typically provide interfaces at the application layer of the OSI-model, but do not replace the hardware layer \cite{dtdef-barbie}. This gap is crucial since the connection between the embedded control system and the actual hardware is a common source of errors.

A solution to this problem is the \textit{Digital Twin Prototype}, which also includes the interfaces and is defined as follows \cite{dtdef-barbie}:

\begin{tcolorbox}
\begin{definition}[Digital Twin Prototype]
A Digital Twin Prototype (DTP) is the software prototype of a \pt. The configurations are equal, yet the connected sensors/actuators are emulated. To simulate the behavior of the \pt, the emulators use existing recordings of sensors and actuators. For continuous integration testing, the DTP can be connected to its corresponding \dt, without the availability of the \pt.
\end{definition}
\end{tcolorbox}

The essence of the \dtp approach lies in replacing all sensors and actuators with emulators or simulations, thereby virtualizing the hardware interfaces. This is key to the implementation. For open standards, various tools enable the creation of virtual interfaces. kernel-level tools and emulators are available for many communication protocols. 
However, this might not be the case for proprietary hardware interfaces. 

\begin{figure}[b]
    \centering
    \includegraphics[width=0.45\textwidth]{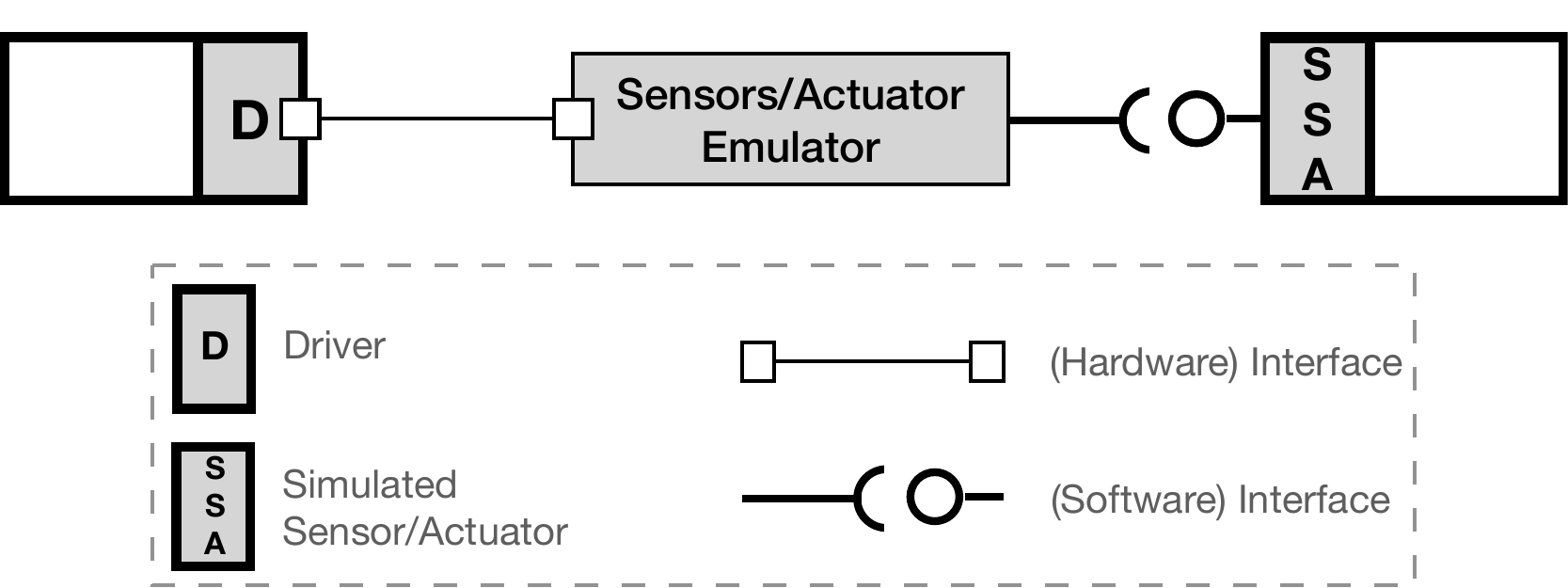}
    \caption{An emulator helps to virtualize a hardware interface and return virtual context from a simulation.}
    \label{fig:emulatorinterfaces}
\end{figure}

In \Cref{fig:emulatorinterfaces}, the general idea of a \dtp setup with interfaces is shown. The approach involves a device driver for a physical object using a virtual hardware interface to connect to an emulated sensor or actuator, which could either use recorded real-world data or rely on simulations. 
The device driver, in this scenario, does not need to distinguish between an actual sensor/actuator and its emulated counterpart.

For the PiCar, the main interfaces, GPIO and I2C, are emulated using Linux kernel tools. The virtual GPIO interaction module (gpio-mockup) and the I2C chip (I2C-stub) are integrated into the container for emulation purposes. Additionally, with Windows 10 and later versions offering a built-in Linux virtual machine through WSL2, these interfaces can also be enabled and used on Windows systems. This setup allows for a flexible and adaptable environment for emulating the PiCar's hardware interactions.

The DC motors use GPIO pins for direction control, and the speed and angle settings for the DC motors and servo motor, respectively, are managed via an I2C interface. The simulation harnesses the ``ros\_control'' package for emulating sensors and actuators. The servo motor emulator reads pulse widths from an I2C register, converting them back into angles, while the DC motor emulator continuously checks its GPIO pin and I2C register to determine direction and speed. Since the PiCar-X's motors are identical, the emulator are identical, too. Hence, the directional handling has to be managed in the driver to ensure coordinated movement. Otherwise, the wheels of the PiCar-X would turn to different directions.

The Ackermann steering \cite{Veneri2020} node activates upon receiving a command. It then sends messages to the motor drivers, setting the speed, and to the servo driver, specifying an angle. The emulators, interfacing with the I2C chip, read these values and publish corresponding commands to the simulated actuators via ROS messages. For the motor emulators, the GPIO pin value determines the direction of motion, with 0 indicating forward and 1 indicating backward.

\begin{figure*}[t]
    \centering
    \includegraphics[width=0.85\textwidth]{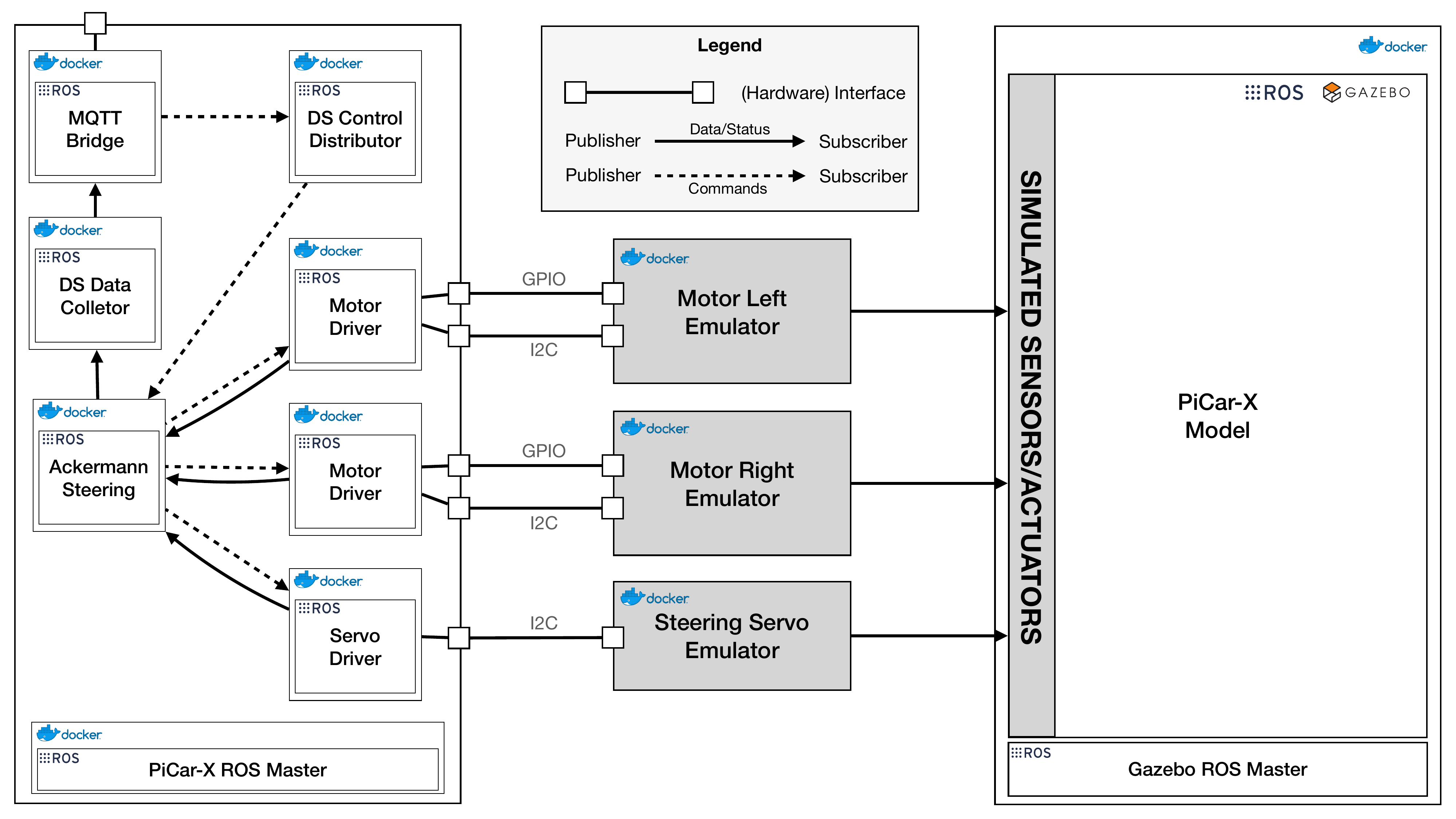}
    \caption{PiCar-X \dtp setup.}
    \label{fig:dtpsetup}
\end{figure*}

The \dtp's setup, as depicted in \Cref{fig:dtpsetup}, effectively separates the \gazebo simulation and the \pt's software through the use of distinct ROS Masters. 
Emulators, connected to the simulation, interface solely through emulated hardware interfaces, making the \pt's software unable to distinguish between actual hardware and simulation. This design enables comprehensive testing and development of the PiCar-X software without requiring the physical hardware, offering a versatile and efficient development environment. Additionally, it introduces a sustainability perspective to the development process \cite{Heithoff2023}. 
This setup has already been utilized in a real-world context, specifically in a smart farming application~\cite{silagecontrol}. Therefore, the PiCar-X example can also be used to explore this approach.

\section{Digital Twin Prototypes in CI/CD-Pipelines}\label{sec:experiment}
The PiCar-X lacks a GPS sensor or equivalent for position determination, and the specifications of its motors, such as torque, are unknown, preventing accurate calculation of its speed. For example, the maximum motor operation corresponds to an I2C register value of 4095, equating to a voltage usage of 3 to 6V. However, tests with the manufacturer's code revealed inconsistencies in motor movement and servo rotation. Without a reliable ground truth, verifying the implementation was challenging.

In the \gazebo simulation for the PiCar-X digital model, these uncertainties, along with incomplete manufacturer information and the model's limitations, hindered accurate velocity calculation for the motor joints. The simulated PiCar-X moved slower when using the I2C value for motor speed, necessitating a speed adjustment factor, which we determined through experimentation.

\begin{figure}[t]
    \centering
    \includegraphics[width=.45\textwidth]{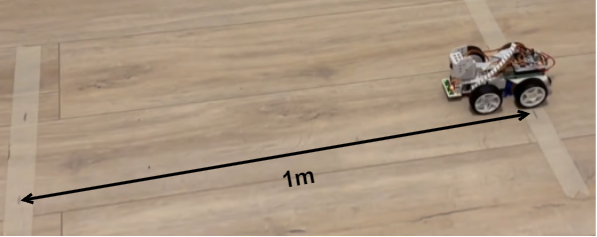}
    \caption{Experiment to determine the PiCar's speed for 1 meter.}
    \label{fig:picarexperiment}
\end{figure}

To calculate the velocity factor for the PiCar-X's digital model motor joints, we first measured the time it took for the physical PiCar-X to travel one meter. This was achieved by positioning the PiCar-X at a starting point and then triggering a script that would drive it at full speed for a set duration, initially one second, see \Cref{fig:picarexperiment}. If the PiCar-X stopped at the one-meter mark within this time frame, the experiment was repeated ten times with the same settings to ensure consistency. We discovered that the PiCar-X typically required about 1.45 seconds to travel one meter, with minor deviations of one to two centimeters from this distance. This variation might be attributed to ROS's non-real-time operation, as even a slight delay (e.g., 50ms) can affect the distance traveled.

\begin{figure*}[b]
    \centering
    \includegraphics[width=\textwidth]{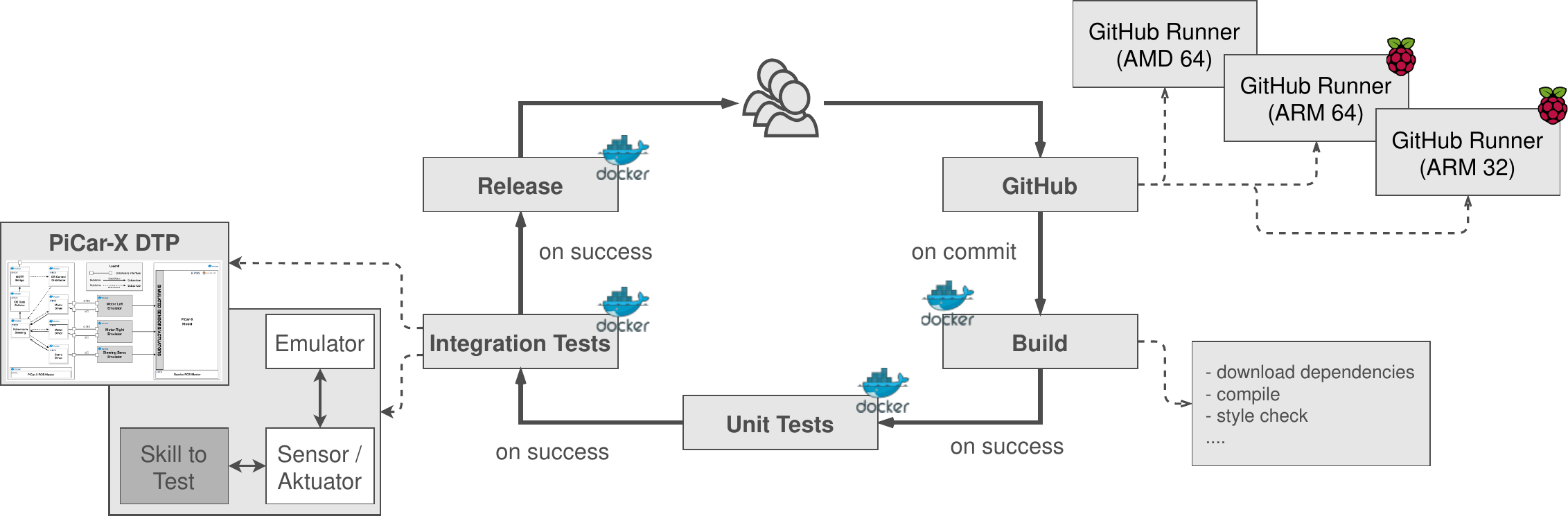} 
    \caption{The DTP can be used for integration testing in CI/CD-Pipelines. An example for the PiCar-X is provided using GitHub Actions.}
    \label{fig:picarxcicd}
\end{figure*}

In the simulation, this procedure was inverted. The PiCar-X moved for 1.45 seconds and the velocity factor was adjusted until a satisfactory approximation was found. Afterwards, the experiment was repeated ten times using this factor to validate the correctness. The PiCar's travel distance in the simulation was monitored by subscribing to a topic that reports the position of the rear wheels. The goal was for the PiCar's right rear wheel, starting at the position $(x,y,z) = (0,0,0)$, to end close to the position $(1,0,0)$.
In the digital model, the deviation in distance was more significant, ranging from two to five centimeters. This larger variance might be due to the additional layer in the message process. In the simulation, the process begins with sending a ROS message to the device driver. This driver then writes values to the emulated I2C chip. Following this, the motor emulator reads these values from the I2C register and subsequently publishes a ROS message to \gazebo, completing the sequence.

The experiment can be repeated and even automated using the simulation. This can be useful in cases where the model is adjusted, for example, when additional components are added. These components can, for instance, affect the PiCar's overall weight, causing the speed in the simulation to differ from the actual speed. In this case, the velocity factors of the motor joints would need to be adjusted. Automated tests would indicate this issue immediately after the changes are committed.

\textcite{studyembeddedtesting} found that test automation is one of the most prominent subjects in testing embedded software. Yet, automatic quality assurance remains a significant challenge, primarily due to the involvement of hardware in the testing process. The predominant testing technique still involves using real systems; though, simulated systems, such as Model-in-the-Loop (MiL), Hardware-in-the-Loop , Software-in-the-Loop (SiL), or Processor-in-the-Loop (PiL), are also widely employed \cite{Bringmann2008, studyembeddedtesting, silagecontrol}. Testing on the actual system necessitates a permanent connection to the testing environment, which can be costly and impractical, particularly for small and medium-sized enterprises \cite{silagecontrol}. This is the reason why the \dtp focuses on being independent from the real system while still supporting the testing of real embedded software.

In order to exemplify the process, we have incorporated the script we used for speed measurement into a CI/CD environment in our GitHub project \cite{abarbiegithub} using GitHub Actions. \Cref{fig:picarxcicd} shows these CI/CD pipelines. Each time a user commits changes, a GitHub Runner is triggered, which builds a Docker container. All dependencies are loaded into this container, and the code gets compiled. This build step could also include style checks to further analyze the quality of the program code. Once the containers are successfully built, unit tests are executed on the module to be tested. If they pass, we move on to the exciting step: the integration tests.

Integration tests, which examine the interaction between sensors/actuators and the control logic, usually require the actual hardware. If this hardware is not available, the tests often cannot be performed. To circumvent this problem, we developed the \dtp in the ARCHES~\cite{dtdef-barbie} project. As presented in \Cref{subsec:dtp}, the \dtp uses emulators instead of real hardware, enabling automated tests of the entire control logic. Our PiCar-X example demonstrates the integration test for speed determination using the virtual context provided by the simulation. After the integration tests are completed successfully, the Docker containers are published.

The presented example already includes PiL, SiL, and MiL tests. To exemplify the PiL tests, we execute the tests on three different processor architectures using three different GitHub Runners: an AMD 64 processor representing the common environment on laptops and desktop workstations, a Raspberry Pi 4B with an ARM64 processor architecture, and a Raspberry 3B+ with an ARM32 processor architecture. The PiCar-X is built around a Raspberry Pi. The SiL tests are exemplified by fully replacing the hardware with emulators. MiL tests are also included, as a simulation is used to provide the virtual context when the PiCar moves. Note that the \gazebo simulation only works on machines with AMD64 chip architectures. To enable integration tests that require virtual context from the simulation, \gazebo must be started on a server using an AMD64 chip architecture, and the ROS nodes running on the Raspberry Pi must connect to this server. ROS itself supports ARM chip architectures.

This CI/CD pipeline setup was also already used in the ARCHES project to automatically test the ocean observation systems. However, our example already utilizes the entire DTP of the PiCar-X. In a professional environment, the individual components would have their own CI/CD pipelines. In those pipelines, only tests relevant to each specific component would be executed. The PiCar example shows how to use virtual interfaces for GPIO and I2C interfaces. Our GitHub repository~\cite{abarbiegithub} also includes examples for RS232 interfaces. The git repository of the ADTF~\cite{ADTF} also includes integration tests for the digital thread.

\section{Conclusion}\label{sec:Conclusion}
We have demonstrated how the different parts of our \dt concept, which we formally specified using the Object-Z notation in a previous publication \textcite{dtdef-barbie}, can be implemented for a PiCar-X by SunFounder. The \pt was implemented using the ADTF~\cite{demomission}. All services are encapsulated in Docker containers. The digital model uses a ROS node connected to a \gazebo simulation. The simulation includes the CAD model of PiCar model, yet does not fully represent the PiCar-X. Bundling the digital model with the \pt software and documentation assets, e.g. blue prints of the different components, form the digital template.

Data/status messages from the \pt are automatically sent to the \ds via a digital thread. The \ds is used to monitor the \pt and all received messages are sent to the digital model. By including an additional node that can be used to manipulate the digital model and that automatically sends the command to the \pt, the \ds becomes a \dt with fully automated messages exchange between \pt and \dt in both directions.

A \dtp of a PiCar-X can be employed by software engineers for embedded software development without a connection to the physical PiCar-X. This also enables automated Software-in-the-Loop testing in CI/CD pipelines \cite{demomission}. In the development phase of a PiCar-X, \dtps pave the way for a paradigm switch from traditional embedded software development processes, e.g. the V-Model, to agile and incremental workflows with CI/CD. This way, \dtps allow a rapid prototyping of embedded software systems and the integration of feedback by stakeholders at a very early development stage. This is not relevant for a PiCar-X, but in larger industrial contexts it is. In automated testing, \dtps act as proxies of the \pts; thus, enabling automatic integration testing of embedded software. The digital thread between a \pt and its digital counterpart is verified via appropriate integration tests with the \dtp.

All in all, the \dtp approach supports engineers to increases the quality of embedded software systems and helps to reduce costs while increasing development speed, which has been reported by both Ebert~\cite{SIsoftwarequality} and Ozkaya~\cite{50yearsse} as reasons for the struggle to achieve quality along with managing costs and efficiency. 
For independent reproduction, the code in published open source on GitHub \cite{abarbiegithub}. To start the \pt, the digital model, the digital shadow, the \dt, and \dtp, different Docker compose files are provided.

\printbibliography

\end{document}